\documentclass[preprint,showpacs,preprintnumbers,amsmath,amssymb,endfloa
ts*]{revtex4}
\usepackage{graphicx}

\begin{document}

\thispagestyle{empty}

\title{Comparison of the hydrodynamic and Dirac models
of the dispersion interaction between graphene and H,
He${}^{\ast}$, or Na atoms}

\author{ Yu.~V.~Churkin, A.~B.~Fedortsov, G.~L.~Klimchitskaya,
and V.~A.~Yurova}

\affiliation{
North-West Technical University,
Millionnaya Street 5, St.Petersburg,
191065, Russia
}

\begin{abstract}
The van der Waals and Casimir-Polder interaction of different atoms
with graphene is investigated using the Dirac model which assumes that
the energy of quasiparticles is linear with respect to the momentum.
The obtained results for the van der Waals coefficients of hydrogen
atoms and molecules and atoms of metastable He${}^{\ast}$ and Na as a
function of separation are compared with respective results found
using the hydrodynamic model of graphene. It is shown that, regardless
of the value of the gap parameter, the Dirac model leads to much
smaller values of the van der Waals coefficients than the hydrodynamic
model. The experiment on quantum reflection of metastable He${}^{\ast}$
and Na atoms on graphene is proposed which is capable to discriminate
between the two models of the electronic structure of graphene. In this
respect the parameters of the phenomenological potential for both
these atoms interacting with graphene described by different models
are determined.
\end{abstract}
\pacs{73.22.Pr, 78.67.-n,  12.20.Ds}
\maketitle

\section{Introduction}

It is common knowledge that dispersion force between two neutral
atoms or molecules arises in second order perturbation theory
from the dipole-dipole interaction \cite{1}. This force takes its
name from the fact that it is caused by the dispersions of dipole
operators, i.e., by quantum fluctuations. At short separations,
where the electromagnetic interaction can be considered as
instantaneous, dispersion force depends only on one fundamental
constant, the Planck constant $\hbar$. In this case it is
referred to as the van der Waals force \cite{2}.
At larger separations, where the retardation of the electromagnetic
interaction becomes significant, dispersion force depends on both
$\hbar$ and the velocity of light $c$. In this case the force is
usually called the Casimir-Polder force \cite{38}.
As a result of interatomic interactions, dispersion forces act also
between an atom (molecule) and a macroscopic body and between two
closely spaced macroscopic bodies (in the retarded regime, the force
between two macrobodies is called the Casimir force \cite{3}).

The van der Waals and Casimir-Polder forces between atoms (molecules)
and material walls play an important role in different physical,
chemical and biological phenomena \cite{1,2}.
The magnitude and the distance dependence of these forces were
measured in several experiments \cite{4}. In the last few years
atom-wall interaction is being studied intensively in connection with
the phenomenon of quantum-reflection \cite{5,6,7,8,9,10,11,11a}.
Theoretical description of atom-wall interaction is given by the
Lifshitz theory \cite{1,2,3,4,12} which expresses the van der Waals
and Casimir-Polder energy and force through the dynamic electric
polarizability of an atom and the frequency-dependent dielectric
permittivity of wall material. A large body of research of atom-wall
interactions for different atoms and wall materials on the basis of
the Lifshitz theory has been performed in the
literature \cite{1,2,3,4,13,14,15,16,17,18,19}.

Considerable recent attention was attracted to two-dimensional carbon
nanostructures, such as graphene, carbon nanotubes and
fullerenes \cite{20}. Carbon nanostructures possess unique
mechanical, electrical and optical properties which are of major
fundamental and applied interest. One of potential applications of
carbon nanostructures is to the problem of hydrogen storage \cite{21}.
Keeping in mind that one-atom-thick nanostructures are not characterized
by the dielectric permittivity, the Lifshitz theory seems to be not
immediately applicable. Because of this, dispersion interaction between
hydrogen atoms (molecules) and a graphite sheet (graphene) or
single-walled carbon nanotubes was investigated mainly by using the
phenomenological density functional theory \cite{22,23,24,25}.
The second-order  perturbation theory was also used \cite{26a}
to calculate line shifts of a two-level atom interacting with
a nanotube.
For the multiwalled carbon nanotubes with at least several walls,
the concept of dielectric permittivity of graphite was shown to be
applicable \cite{26}.

The application of the Lifshitz theory of dispersion interaction to
one-atom-thick carbon nanostructures calls for
introduction of some other quantity which determines the
reflection coefficients instead of usually used
dielectric permittivity.
In order to extend the Lifshitz theory to the case of carbon systems,
the description of graphene in terms of the two-dimensional
free electron gas \cite{27} was used. Graphene was considered as an
infinitesimally thin positively charged flat sheet, carring a
homogeneous fluid with some mass and negative charge densities (the
hydrodynamic model). The interaction of the electromagnetic oscillations
with such a sheet was described by the special reflection
coefficients \cite{28,29}. The proposed approximate description was
applied for the calculation of dispersion interaction between two
parallel graphene sheets \cite{30}, between graphene and material
plate \cite{31}, graphene and an atom or a molecule, and between an atom
or a molecule and a single-walled carbon nanotube \cite{32}.
The hydrodynamic model, however, does not take into account that
low-energy excitations in graphene are massless Dirac fermions except
for the fact that they move with a Fermi velocity rather than with the
speed of light \cite{20,33}. As a consequence, at low energies the
dispersion relation for quasiparticles in graphene is approximately
linear with respect to the momentum measured relatively to the corners
of the graphene Brillouin zone. Under the assumption that these
properties, which are referred to as the Dirac model, hold at any energy,
the reflection coefficients for the electromagnetic oscillations on
graphene were found \cite{34}. These coefficients were used to calculate
the Casimir interaction between a graphene and a parallel ideal metal
plane \cite{34}.

In this paper we investigate the dispersion interaction of atoms
(molecules) with graphene described by the Dirac model. We consider
an atom and a molecule of hydrogen, and also atoms of metastable
He${}^{\ast}$ and Na often used in experiments on quantum reflection.
The van der Waals coefficient $C_3$ is computed by using the Lifshitz
formula as a function of atom-graphene separation. The obtained results are
compared with those obtained previously using the hydrodynamic model of
the electronic structure of
graphene. It is shown that in all cases the magnitudes of the
van der Waals and Casimir-Polder energy computed with the help of the Dirac
model are less than with the hydrodynamic model. Differences between the
two sets of results increase with the increase of separation.
We also find the dependence of the predicted van der Waals and
Casimir-Polder energies on the value of the gap parameter.
We propose the experiment allowing to discriminate between the predictions
of the Dirac and hydrodynamic models of graphene. For this purpose the
parameters of phenomenological potential of atom-graphene interaction
are determined for atoms of  metastable He${}^{\ast}$ and Na.

The paper is organized as follows. In Sec.~II we briefly present the
Lifshitz formula for atom-wall interaction and the forms of reflection
coefficients in both models. Section~III contains computational results
for the van der Waals coefficient $C_3$ as a function of separation and
gap parameter for atoms and molecules of hydrogen. In Sec.~IV similar
results are presented for  atoms of  metastable He${}^{\ast}$ and Na.
The parameters of phenomenological potential for these atoms are also
determined for the needs of proposed experiment. In Sec.~V the reader
will find our conclusions and discussion.

\section{Two models for interaction of atoms with graphene}

The van der Waals and Casimir-Polder energy in the configuration of
a microparticle (an atom or a molecule) and a  plane structure can be
expressed in terms of reflection coefficients of the electromagnetic
oscillations on this structure in the following way \cite{3,12}
\begin{eqnarray}
&&
E(a)=-\frac{\hbar}{2\pi}\int_{0}^{\infty}\!\!d\xi
\alpha(i\xi)
\int_{0}^{\infty}k_{\bot}dk_{\bot}\,q\,e^{-2aq}
\nonumber \\
&&~~
\times
\left\{2r_{\rm TM}(i\xi,k_{\bot})-\frac{\xi^2}{q^2c^2}\left[
r_{\rm TM}(i\xi,k_{\bot})+r_{\rm TE}(i\xi,k_{\bot})\right]\right\}.
\label{eq1}
\end{eqnarray}
\noindent
Here, $a$ is the separation distance between the microparticle and
the plane, $k_{\bot}$ is the magnitude of the wave vector projection on the
plane (the latter is perpendicular to the $z$ axis),
$\alpha(i\xi)$ is the dynamic polarizability of a microparticle
calculated along the imaginary frequency axis, and
$q^2=k_{\bot}^2+\xi^2/c^2$.
The reflection coefficients for two independent
polarizations of the electromagnetic field, transverse magnetic and
transverse electric, $r_{\rm TM,TE}$,
are also calculated along the imaginary frequency axis.
In the framework of the scattering approach (see, for instance,
Refs.~\cite{3,35a}) it can be proved that under some conditions
Eq.~(\ref{eq1}) holds for any planar structure with appropriately
chosen reflection coefficients.
Specifically, the use of Eq.~(\ref{eq1}) is justified
if a plane structure (a material of
the plate or a graphene) is in thermal equilibrium with an
environment at some not too high temperature, and an atom is in
the ground state \cite{13}. For instance, the probabilities
that an atom
is in an excited state are distributed according to the Boltzmann
law and are negligibly small at room temperature ($T=300\,$K).
As a result,
 the zero-temperature Lifshitz formula (\ref{eq1})
 can be used  at
sufficiently small separations (see below for quantitative
evaluation of its application region).

There are different models of reflection coefficients for
graphene using different boundary conditions.
 In the framework of the hydrodynamic model of
the electronic structure of graphene they can
be presented in the form \cite{28,29,30,31,32}
\begin{eqnarray}
&&
r_{\rm TM}(i\xi,k_{\bot})\equiv r_{\rm TM}^{(h)}(i\xi,k_{\bot})
=\frac{c^2qK}{c^2qK+\xi^2},
\nonumber \\
&&
r_{\rm TE}(i\xi,k_{\bot})\equiv r_{\rm TE}^{(h)}(i\xi,k_{\bot})
=-\frac{K}{K+q},
\label{eq2}
\end{eqnarray}
\noindent
where the wave number of the graphene sheet
$K=6.75\times 10^{5}\,\mbox{m}^{-1}$ corresponds to the frequency
$\omega_{K}=cK=2.02\times 10^{14}\,$rad/s.
As was explained in Sec.~I, the hydrodynamic model does not take into
account some important properties of graphene which hold at low energies,
specifically that the energy of quasiparticles is a linear function of
the momentum.
These properties are well described by the Dirac model
of the electronic structure of graphene where the
reflection coefficients are given by \cite{34}
\begin{eqnarray}
&&
r_{\rm TM}(i\xi,k_{\bot})\equiv r_{\rm TM}^{(D)}(i\xi,k_{\bot})
=\frac{\alpha q\Phi(\tilde{q})}{2\tilde{q}^2+\alpha q\Phi(\tilde{q})},
\nonumber \\
&&
r_{\rm TE}(i\xi,k_{\bot})\equiv r_{\rm TE}^{(D)}(i\xi,k_{\bot})
=-\frac{\alpha\Phi(\tilde{q})}{2q+\alpha\Phi(\tilde{q})}.
\label{eq3}
\end{eqnarray}
\noindent
Here $\alpha=e^2/(\hbar c)\approx1/137$ is the fine-structure constant,
$\tilde{q}^2=(v_{\rm F}^2k_{\bot}^2+\xi^2)/c^2$,
$v_{\rm F}\approx 10^{6}\,$m/s is the Fermi velocity, and the function
$\Phi$ determines the polarization tensor in an external electromagnetic
field in the one-loop approximation in three dimensional space-time.
The explicit form of this function along the imaginary frequency axis
is the following \cite{34}
\begin{equation}
\Phi(\tilde{q})=N\left(\tilde\Delta+
\frac{\tilde{q}^2-4\tilde\Delta^2}{2\tilde{q}}\,
\arctan\frac{\tilde{q}}{2\tilde\Delta}\right),
\label{eq4}
\end{equation}
\noindent
where for graphene $N=4$, $\tilde\Delta=\Delta/(\hbar c)$, and the
value of the gap parameter $\Delta$ is not well known. The upper bound
on $\Delta$ is approximately equal to 0.1\,eV, but it might be also
much smaller \cite{20,34}.

Below, we perform computations of the van der Waals and Casimir-Polder
interaction energy between different atoms and graphene. For this
purpose it is useful to introduce the dimensionless variable
$y=2qa$ instead of $k_{\bot}$.
Then Eq.~(\ref{eq1}) can be identically rewritten in the form
\begin{equation}
E(a)=-\frac{C_3(a)}{a^3},
\label{eq5}
\end{equation}
\noindent
where the van der Waals coefficient $C_3(a)$ is given by
\begin{eqnarray}
&&
C_3(a)=\frac{\hbar}{16\pi}
\int_{0}^{\infty}\!dye^{-y}
\int_{0}^{cy/(2a)}\!\!d\xi
\alpha(i\xi)
\nonumber \\
&&~~
\times
\left\{2y^2r_{\rm TM}(i\xi,y)-\frac{4a^2\xi^2}{c^2}\left[
r_{\rm TM}(i\xi,y)+r_{\rm TE}(i\xi,y)\right]\right\}.
\label{eq6}
\end{eqnarray}
\noindent
Either the hydrodynamic- or Dirac-model
reflection coefficients (\ref{eq2}) and (\ref{eq3}), respectively,
can be substituted here
keeping in mind that in terms of new
variable it holds
\begin{equation}
q=\frac{y}{2a},\quad
\tilde{q}=\left[\frac{v_{\rm F}^2}{c^2}\frac{y^2}{4a^2}+
\left(1-\frac{v_{\rm F}^2}{c^2}\right)\frac{\xi^2}{c^2}
\right]^{1/2}.
\label{eq7}
\end{equation}
\noindent
Notice that Eq.~(\ref{eq5}) with constant $C_3$ presents the energy
of nonretarded van der Waals interaction between an atom and a wall
which is valid only at small separations  of about
2--3\,nm\cite{1,2,3,4,12,13,14}
(at shorter separations the description of graphene by means of
the boundary conditions becomes inapplicable and one should take
into account an atomic structure of the sheet).
If, however, $C_3$ depends on separation in accordance with Eq.~(\ref{eq6}),
the interaction energy (\ref{eq5}) includes both the nonretarded and
retarded (relativistic) regimes.
 As a result, Eqs.~(\ref{eq5}) and (\ref{eq6}) are
applicable up to separation distances where the thermal effects become
significant. It is well known \cite{3,13} that at room temperature
the zero-temperature Lifshitz formula provides very exact computational
results of atom-plate interaction up to separations of about
$1\,\mu$m (see Sec.~IV for a computational example).
As to the function $\Phi$ determining the polarization tensor,
relativistic thermal quantum field theory does not predict
a noticeable change of this quantity at room temperature, as
compared to the case of zero temperature (note that computation
in Ref.~\cite{38a} was, strictly speaking, performed in a
nonretarded regime).

\section{Interaction of hydrogen atoms and molecules with graphene}

Here, we compare the van der Waals coefficients for interaction of
hydrogen atoms and molecules with graphene in the cases when computations
are performed with the help of Dirac and hydrodynamic models.
To perform computations using Eq.~(\ref{eq6}), one needs some expressions
for the atomic and molecular dynamic polarizabilities of hydrogen.
As was shown in Refs.~\cite{14,26}, for the calculation of dispersion
forces, the polarizabilities can be represented with sufficient
precision in the framework of single-oscillator model
\begin{equation}
\alpha(i\xi)=\frac{\alpha(0)}{1+\frac{\xi^2}{\omega_0^2}},
\label{eq8}
\end{equation}
\noindent
where $\alpha(0)$ is the static polarizability and $\omega_0$ is the
characteristic frequency. For a hydrogen atom it was found \cite{35}
that $\alpha(0)\equiv\alpha_a(0)=4.50\,$a.u. and
$\omega_0\equiv\omega_{0a}=11.65\,$eV (remind that 1\,a.u.
of polarizability
is equal to $1.482\times 10^{-31}\,\mbox{m}^{3}$).
For a hydrogen molecule it holds that \cite{35}
$\alpha(0)\equiv\alpha_m(0)=5.439\,$a.u. and
$\omega_0\equiv\omega_{0m}=14.09\,$eV.

In Fig.~1(a) we present the computational results for the van der Waals
coefficient $C_{3,\rm H}$ (in atomic units)
for a hydrogen atom interacting with graphene
as a function of separation.
Note that one atomic unit for $C_3$ is equal to
$4.032\times 10^{-3}\,\mbox{eV\,nm}^{3}$.
Computations were performed using
Eqs.~(\ref{eq6}) and (\ref{eq8}) with the hydrodynamic-model reflection
coefficients (\ref{eq2}) (the dashed line) and with the Dirac-model
reflection coefficients (\ref{eq3}) (the upper and lower solid lines
obtained with the lower and upper bounds for graphene gap parameter
$\Delta$, respectively). As an upper bound, the value 0.1\,eV was
chosen (see Sec.~II). Keeping in mind that below some $\Delta_{\min}$
the values of $C_3$ are insensitive to further decrese of $\Delta$
(see below), the value of $10^{-15}\,$eV was chosen as a lower bound.
As is seen in Fig.~1(a), the values of $C_{3,\rm H}$ computed using
the hydrodynamic and Dirac models are significantly different and the
ratio $C_{3,\rm H}^{(h)}/C_{3,\rm H}^{(D)}$ increases with the increase
of separation distance. Thus, at the shortest separation $a=3\,$nm
it holds $C_{3,\rm H}^{(h)}/C_{3,\rm H}^{(D)}=1.065$. When separation varies
to 5, 10, 20, 50, and 100\,nm this ratio increases to 1.19, 1.44, 1.85,
2.85, and 4.21, respectively (we have used the computational data
related to the lower solid line with $\Delta=0.1\,$eV).
At large separations of about 100\,nm both $C_{3,\rm H}^{(h)}(a)$
and $C_{3,\rm H}^{(D)}(a)$ decrease with separation as $1/a$
leading to $E_{\rm H}^{(h,D)}(a)\sim 1/a^4$. This is a typical
behavior of the atom-plate interaction at relativistic separations
\cite{38,3,4} which holds also for other atomic systems considered
below. In the nonretarded region $a<3\,$nm the Casimir-Polder
energy depends on separation as $a^{-7/2}$. This region is, however,
outside the application region of our formalism based on the use
of boundary conditions.

The relationship between the two solid lines in Fig.~1(a) demonstrates
that the influence of the gap parameter on the value of $C_{3,\rm H}$
is rather moderate. This is illustrated in more detail in Fig.~1(b)
where $C_{3,\rm H}$ is plotted as a function of $\log_{10}\Delta$.
The solid lines labeled 1, 2, and 3 are computed at separations
$a=5$, 50, and 100\,nm, respectively. For the line labeled 1 the relative
difference between the maximum and minimum values of $C_{3,\rm H}$,
$(C_{3,\rm H}^{\max}-C_{3,\rm H}^{\min})/C_{3,\rm H}^{\min}=6.6$\%.
For the lines labeled 2 and 3 this relative difference is equal to
16.4\% and 31.3\%, respectively. For sufficiently small $\Delta_{\min}$
(different at each separation) $C_{3,\rm H}$ becomes constant and does not
depend on further decrease of $\Delta$. Thus, for lines labeled 1, 2, and 3
the van der Waals coefficient is constant for gap parameters
satisfying the inequalities $\Delta\leq 0.01\,$eV, $\Delta\leq 0.004\,$eV,
and $\Delta\leq 0.001\,$eV, respectively.

The computational results for a hydrogen molecule differ from the
respective results for a hydrogen atom quantitatively but not
qualitatively. In Fig.~2 the van der Waals coefficient
$C_{3,{\rm H}_2}$ is plotted as a function of separation [the dashed line
is computed using the hydrodynamic model and the two solid lines
using the Dirac model with the same gap parameters as in Fig.~1(a)].
Here, at the shortest separation $a=3\,$nm it holds
$C_{3,{\rm H}_2}^{(h)}/C_{3,{\rm H}_2}^{(D)}=1.045$, i.e., the
computational results are slightly closer than for a hydrogen atom.
With the increase of separation to 5, 10, 20, 50, and 100\,nm the ratio
of the van der Waals coefficients computed using the two models
increases to 1.18, 1.45, 1.89, 3.00, and 4.63, respectively, i.e.,
becomes larger than respective ratio for a hydrogen atom. As to the role
of the gap parameter of graphene, its value only slightly influences
the computational results. Figures 1(a) and 2 raise a question if it is
possible to discriminate between the two models of graphene from
experiments on quantum reflection. This question is considered in the
next section.

\section{Quantum reflection of metastable H$\mbox{e}{}^{\ast}$ and
N$\mbox{a}$ atoms on graphene}

Experiments on quantum reflection \cite{5,6,7,8,9,10,11,11a} of cold
atoms on different surfaces demonstrated that the reflection amplitude
depends critically on the form of interaction potential. Because of this,
using the experimental data for reflection amplitudes, it is possible to
restore the form of potential within the limits of experimental error.
If the two models of graphene (the hydrodynamic one and the Dirac one)
lead to the values of interaction energy which differ dramatically,
it might be possible to experimentally exclude some model or find it
consistent with the data. To make reflection amplitudes more
pronounced it is customary in experiments on quantum reflection to use
atoms with large static electric polarizability. This allows to obtain
more than one order of magnitude larger values of the van der Waals
coefficient $C_3$ when compared to the values of
the same coefficient for
hydrogen atoms and molecules plotted in
Figs.~1(a,b) and 2.

We begin with an atom of metastable He${}^{\ast}$ interacting
with graphene by means of dispersion interaction.
For He${}^{\ast}$ atom it holds \cite{36}
$\alpha(0)\equiv\alpha_{{\rm He}^{\ast}}(0)=315.63\,$a.u. and
$\omega_0\equiv\omega_{0,{\rm He}^{\ast}}=1.18\,$eV.
Computations were performed using
Eqs.~(\ref{eq6}) and (\ref{eq8}) with reflection coefficients, as defined
in the hydrodynamic  (\ref{eq2})
or Dirac (\ref{eq3}) models of graphene.
The computational results for the van der Waals coefficients
$C_{3,{\rm He}^{\ast}}$ as a function of separation are presented in
Fig.~3 by the dashed line (the hydrodynamic model) and by the two solid
lines (the Dirac model) obtained with two values of the gap parameter
of graphene indicated in Sec.~III.
Similar to the case of hydrogen atom and molecule the ratio
$C_{3,{\rm He}^{\ast}}^{(h)}/C_{3,{\rm He}^{\ast}}^{(D)}$ is larger than
unity (it is equal to 1.33 at $a=3\,$nm and increases with the increase of
separation). At $a=5$, 10, 20, 50, and 100\,nm this ratio is equal to
1.47, 1.76, 2.23, 3.33, and 4.78. Thus, it is larger than for atoms and
molecules of hydrogen at all separations. What is more important,
the magnitudes of $C_{3,{\rm He}^{\ast}}^{(D)}$ computed with the Dirac
model are by a factor of 18 (at $a=3\,$nm) and by a factor of 46
(at $a=100\,$nm) larger than respective values of $C_{3,{\rm H}}^{(D)}$
computed using the same model. In a similar way,
$C_{3,{\rm He}^{\ast}}^{(h)}(3\,\mbox{nm})=
23C_{3,{\rm H}}^{(h)}(3\,\mbox{nm})$ and
$C_{3,{\rm He}^{\ast}}^{(h)}(100\,\mbox{nm})=
52C_{3,{\rm H}}^{(h)}(100\,\mbox{nm})$.
This makes experiments on quantum reflection of He${}^{\ast}$ atoms
on graphite feasible.
Note that the role of thermal effects at separations considered is
negligibly small. For example, we have computed the values of
$C_{3,{\rm He}^{\ast}}^{(D)}$ at $a=0.5\,\mu$m both at $T=0$ and
$T=300\,$K using the zero-temperature and the thermal Lifshitz
formulas. In the framework of the Dirac model the following
respective values were obtained:
$C_{3,{\rm He}^{\ast}}^{(D)}=0.0183505\,$a.u. and
$C_{3,{\rm He}^{\ast}}^{(D)}=0.0183565\,$a.u.
This means that at $a=0.5\,\mu$m the relative difference between
the results  obtained using the zero-temperature and thermal
Lifshitz formula is equal to only 0.033\%.

It is interesting also to analyze the influence of a gap parameter of
graphene on the value of $C_{3,{\rm He}^{\ast}}$ illustrated in Fig.~3
by the upper and lower solid lines computed with $\Delta=10^{-15}\,$eV
and with $\Delta=0.1\,$eV, respectively. Thus, at $a=5\,$nm it holds
$(C_{3,{\rm He}^{\ast}}^{\max}-
C_{3,{\rm He}^{\ast}}^{\min})/C_{3,{\rm He}^{\ast}}^{\min}=3.7$\%.
At separations $a=50$ and 100\,nm the same relative difference is
equal to 28\% and 46\%, respectively. From this it follows that at
large separations the impact of the value of a gap parameter $\Delta$ on
$C_{3,{\rm He}^{\ast}}^{(D)}(a)$ is somewhat stronger than on
$C_{3,{\rm H}}^{(D)}(a)$.

Another atom suitable in experiments on quantum reflection on graphene
is Na. For Na atom it holds \cite{37}
$\alpha(0)\equiv\alpha_{{\rm Na}}(0)=162.68\,$a.u. and
$\omega_0\equiv\omega_{0,{\rm Na}}=1.55\,$eV.
The computational results for the van der Waals coefficients
$C_{3,{\rm Na}}^{(h)}$ (the dashed line) and $C_{3,{\rm Na}}^{(D)}$
(the two solid lines for two values of the gap parameter indicated above)
as a function of separation are presented in Fig.~4.
At $a=3\,$nm we obtain  $C_{3,{\rm Na}}^{(h)}/C_{3,{\rm Na}}^{(D)}=1.40$.
With the increase of separation to $a=5$, 10, 20, 50, and 100\,nm
this ratio increases to 1.55, 1.87, 2.39, 3.61, and 5.29, respectively.
The magnitudes of $C_{3,{\rm Na}}^{(D)}$ are by a factor 11.6
(at $a=3\,$nm) and by a factor of 26 (at $a=100\,$nm) larger than the
respective values of $C_{3,{\rm H}}^{(D)}$. In a similar way,
$C_{3,{\rm Na}}^{(h)}(3\,\mbox{nm})=15C_{3,{\rm H}}^{(h)}(3\,\mbox{nm})$
and
$C_{3,{\rm Na}}^{(h)}(100\,\mbox{nm})=33C_{3,{\rm H}}^{(h)}(100\,\mbox{nm})$.
The exceeding of $C_{3,{\rm Na}}^{(h)}$ relative to
$C_{3,{\rm H}}^{(h)}$
is not so large as for He${}^{\ast}$ atoms, but still
sufficient for experimental purposes.

The upper and lower solid lines in Fig.~4 illustrate the influence
of a gap parameter of graphene $\Delta$ on the value of
$C_{3,{\rm Na}}$. Thus, at $a=5\,$nm it holds
$(C_{3,{\rm Na}}^{\max}-C_{3,{\rm Na}}^{\min}/C_{3,{\rm Na}}^{\min}=3.2$\%.
At separations $a=50$ and 100\,nm the relative difference between
maximum and minimum values of $C_{3,{\rm Na}}$ is equal to 25.6\% and
42\%, respectively. These results are similar to the respective
results for He${}^{\ast}$ atoms.

In almost all papers devoted to the experimental investigation
of quantum reflection \cite{5,6,10,11a} the comparison of the
measurement data with theory is performed with the use of a phenomenological
potential for the atom-wall interaction. This simplifies the calculation
of theoretical reflection amplitudes to be compared with the
experimental ones. The most often used phenomenological potential has
the form \cite{3,4,5,6,10,11a}
\begin{equation}
E^{\rm ph}(a)=-\frac{C_4}{a^3(a+l)},
\label{eq9}
\end{equation}
\noindent
where $l$ is a characteristic parameter with the dimension of length
that depends on the material of the plate. It is assumed that at
short separations $a\ll l$ (typically at separations of a few nanometers
where the retardation effects are negligibly small), $E^{\rm ph}(a)$
coincides with the van der Waals interaction energy
between an atom and usual material plate [Eq.~(\ref{eq5}) with
practically constant $C_3$], so that $C_4=lC_3$. At large separations,
where $l$ is negligibly small in comparison with $a$, the
phenomenological potential (\ref{eq9}) coincides with the Casimir-Polder
retarded interaction
energy $-C_4/a^4$ obtained for ideal metal wall \cite{3,38}. The
comparison of computational results obtained using Eq.~(\ref{eq9}) with
measurement data for the reflection amplitudes allows one to determine the
experimental values for the parameters $l$ and $C_4$.
Note that Eq.~(\ref{eq9}) does not agree with a nonretarded asymptotic
limit for atom-graphene interaction (see Sec.~III).
As shown below, Eq.~(\ref{eq9}) provides a very good agreement with computations
using the Lifshitz formula only at moderate separations.
In Ref.~\cite{17} the accuracy of the phenomenological potential
was investigated by the comparison with exact results obtained using
the Lifshitz theory. It was shown that the phenomenological
potential leads to quantitatively correct results only in the
limit of short separation distances below 100\,nm.
This conclusion was confirmed experimentally in Ref.~\cite{40}
where deviations between the theoretical prediction using the
phenomenological potential and the measurement data within the
separation region from 160 to 220\,nm were demonstrated.
Below we use the phenomenological potential (\ref{eq9}) only
at separations below 100\,nm and compare the obtained results with
those using the Lifshitz theory. The separation regions where
Eq.~(\ref{eq9}) leads to less than 1\% error are specified.

Here, we perform the best fit of the phenomenological potential (\ref{eq9})
to the energies $E^{(h)}(a)$ and $E^{(D)}(a)$ computed above for the
interaction of He${}^{\ast}$ and Na atoms with graphene described using
the hydrodynamic and Dirac models, respectively. In doing so we first
use the gap parameter of graphene $\Delta=0.1\,$eV. The results of the
best fit of Eq.~(\ref{eq9}) to the dashed line in Fig.~3 (a He${}^{\ast}$
atom interacting with graphene described by the hydrodynamic model)
are the following:
\begin{equation}
C_{4,{\rm He}^{\ast}}^{(h)}=85.11\,\mbox{a.u.},\qquad
l_{{\rm He}^{\ast}}^{(h)}=72.77\,\mbox{nm}.
\label{eq10}
\end{equation}
\noindent
We note that 1\,a.u. of $C_4$ is equal to
$4.032\times 10^{-3}\,\mbox{eV\,nm}^{4}$.
For the Dirac model, by making the best fit of Eq.~(\ref{eq9}) to the
lower solid line in Fig.~3, we obtain
\begin{equation}
C_{4,{\rm He}^{\ast}}^{(D)}=12.59\,\mbox{a.u.},\qquad
l_{{\rm He}^{\ast}}^{(D)}=11.18\,\mbox{nm}.
\label{eq11}
\end{equation}
\noindent
It can be seen that the value of $E^{\rm ph}(a)$ from Eq.~(\ref{eq9})
with the parameters (\ref{eq10}) deviate from the dashed line in Fig.~3
for less than 1\% in the separation region from 10 to 60\,nm.
At the shortest and largest separations (3 and 100\,nm, respectively)
relative deviations achieve 10\%.
For the Dirac model, Eq.~(\ref{eq9}) with the parameters (\ref{eq11})
deviates from the lower solid line in Fig.~3 for less than 1\% at
separations from 6 to 100\,nm. The largest relative deviation of 5\%
holds at $a=3\,$nm.
So large differences in the parameters of phenomenological potential,
as predicted by the two models of graphene (by a factor of 6.8 for
$C_{4,{\rm He}^{\ast}}$ and by a factor of 6.5 for $l_{{\rm He}^{\ast}}$),
make possible to decide between these models experimentally.
The use of alternative values of the gap parameter leaves this conclusion
unchanged. For example, repeating the same fit with $\Delta=10^{-15}\,$eV
(the upper solid line in Fig.~3) one arrives at
$C_{4,{\rm He}^{\ast}}^{(D)}=18.04\,$a.u. and
$l_{{\rm He}^{\ast}}^{(D)}=18.22\,$nm. These are by factors 4.7 and 4.0
smaller than respective values (\ref{eq10}) obtained using the
hydrodynamic model of graphene.

Similar fit of Eq.~(\ref{eq9}) was performed to the lines of Fig.~4
representing the computational results for atoms of Na.
This leads to
\begin{equation}
C_{4,{\rm Na}}^{(h)}=50.82\,\mbox{a.u.},\qquad
l_{{\rm Na}}^{(h)}=66.92\,\mbox{nm}
\label{eq12}
\end{equation}
\noindent
for the hydrodynamic model and
\begin{equation}
C_{4,{\rm Na}}^{(D)}=7.11\,\mbox{a.u.},\qquad
l_{{\rm Na}}^{(D)}=9.77\,\mbox{nm}
\label{eq13}
\end{equation}
\noindent
for the Dirac model of graphene.
The energy (\ref{eq9})
with parameters (\ref{eq12}) deviates from the dashed line in Fig.~4
for less than 1\% in the separation region from 10 to 60\,nm.
At separations of 3 and 100\,nm the relative deviations achieve 11\%.
For the Dirac model, Eqs.~(\ref{eq9}) and (\ref{eq13}) describe the
lower solid line in Fig.~4 with less than 1\% error in the region from 6
to 100\,nm with maximum relative deviation of 5.7\% at $a=3\,$nm.
Thus, for Na atoms the predictions of the two models of graphene for
$C_{4,{\rm Na}}$ and $l_{{\rm Na}}$ differ by factors of 7.1 and 6.8,
respectively. This makes possible an experimental choice between these
two models. The above computations were performed with $\Delta=0.1\,$eV.
For the extremely small gap parameter $\Delta=10^{-15}\,$eV (the upper
solid line in Fig.~4), one arrives at the values
$C_{4,{\rm Na}}^{(D)}=9.74\,$a.u. and $l_{{\rm Na}}^{(D)}=15.45\,$nm.
These parameters of the phenomenological potential (\ref{eq9}) are
still less than the parameters obtained with the use of hydrodynamic
model by factors of 5.2 and 4.3, respectively.
This confirms that the experimental choice between the two models is
possible with any value of the gap parameter.

\section{Conclusions and discussion}

In the foregoing, we have investigated the van der Waals and
Casimir-Polder interaction of different atoms
with graphene on the basis of fundamental
Lifshitz theory of atom-wall interaction. This was done by using
the two approximate models of the electronic structure of
graphene proposed in the literature,
the hydrodynamic model and the Dirac model. The respective two sets
of reflection coefficients for the electromagnetic oscillations on
graphene were used in computations of the van der Waals coefficients
describing dispersion interaction of hydrogen (H and H${}_2$), and
He${}^{\ast}$ and Na atoms with graphene as a function of separation.
It was shown that the hydrodynamic and Dirac models of graphene lead
to drastically different predictions for the strength of atom-graphene
interaction. This conclusion was obtained independently of the value
of the gap parameter of graphene which was allowed to reduce from
the experimental upper limit to almost zero.

Keeping in mind that both models of graphene are idealizations valid
for different values of parameters, the new experiment was proposed
capable to discriminate between the alternative theoretical
predictions. For this purpose it was suggested to consider the
quantum reflection of He${}^{\ast}$ and Na atoms on graphene.
Due to large static electric polarizabilities, the van der Waals
coefficients for the interaction of
He${}^{\ast}$ and Na atoms with graphene
were shown to be in several tens times greater than for H and H${}_2$.
This makes the observation of quantum reflection of He${}^{\ast}$ and Na
atoms on graphene feasible. For the needs of proposed experiment,
we have determined the parameters of phenomenological potential for
He${}^{\ast}$ and Na atoms interacting with graphene described by
both the hydrodynamic and Dirac models. It was shown that the parameters
obtained using the hydrodynamic model are by a factor varying from 4.0 to
7.1 greater than the parameters obtained using the Dirac model.
This demonstrates that the proposed experiment should be capable to
discriminate between the predictions  of the Dirac and hydrodynamic
models of graphene.

In the future it is supposed to employ the Dirac model for the computation
of dispersion interaction between graphene and material plates made of
different materials and between various atoms and molecules and carbon
nanotubes.

\section*{Acknowledgments}
This work  was supported by the Grant of the Russian Ministry of
Education P--184. G.L.K.\ was also partially supported by
CNPq (Brazil).


\begin{figure*}[h]
\vspace*{-4.cm}
\centerline{
\includegraphics{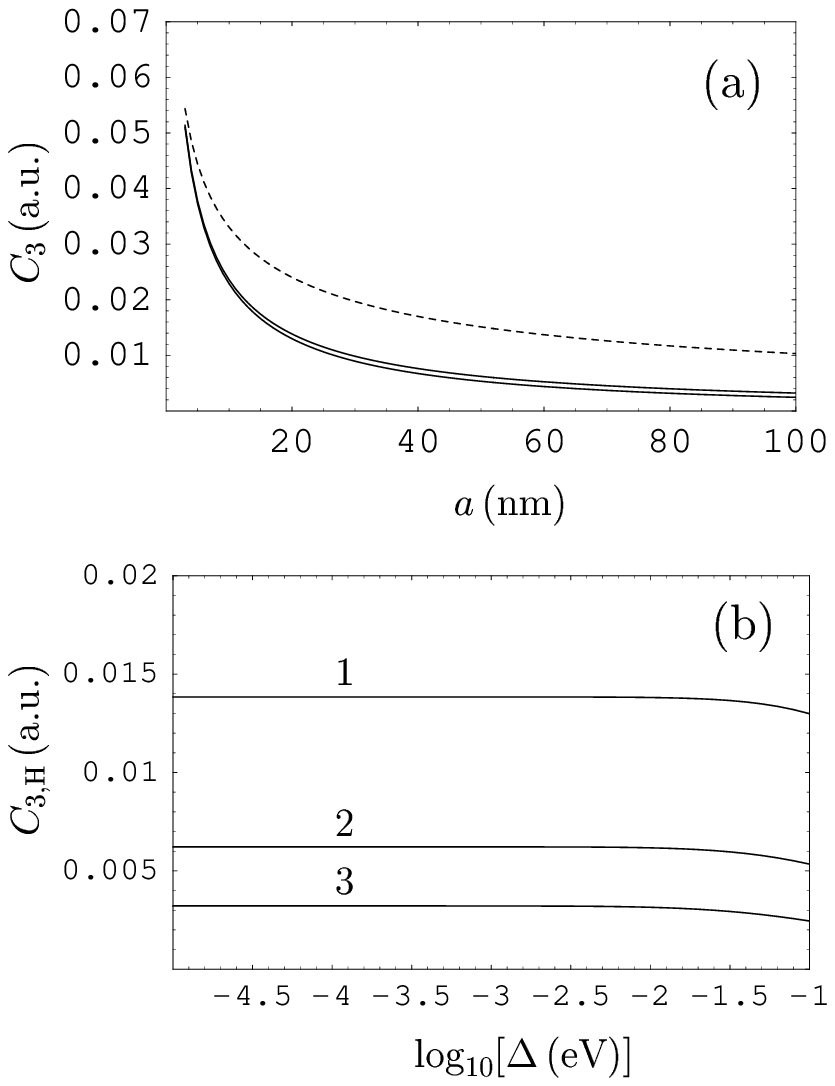}
} \vspace*{-11.cm} \caption{The van der Waals coefficient for a
H atom interacting with graphene as a function of (a) separation
and (b) gap parameter. The dashed and the two solid lines are
computed using the hydrodynamic model and the Dirac model with
two different values of the gap parameter indicated in the text.
The solid lines labeled 1, 2, and 3 are computed using the Dirac
model at separations $a=5$, 50, and 100\,nm, respectively.}
\end{figure*}
\begin{figure*}[h]
\vspace*{-13.cm}
\centerline{\hspace*{2.5cm}
\includegraphics{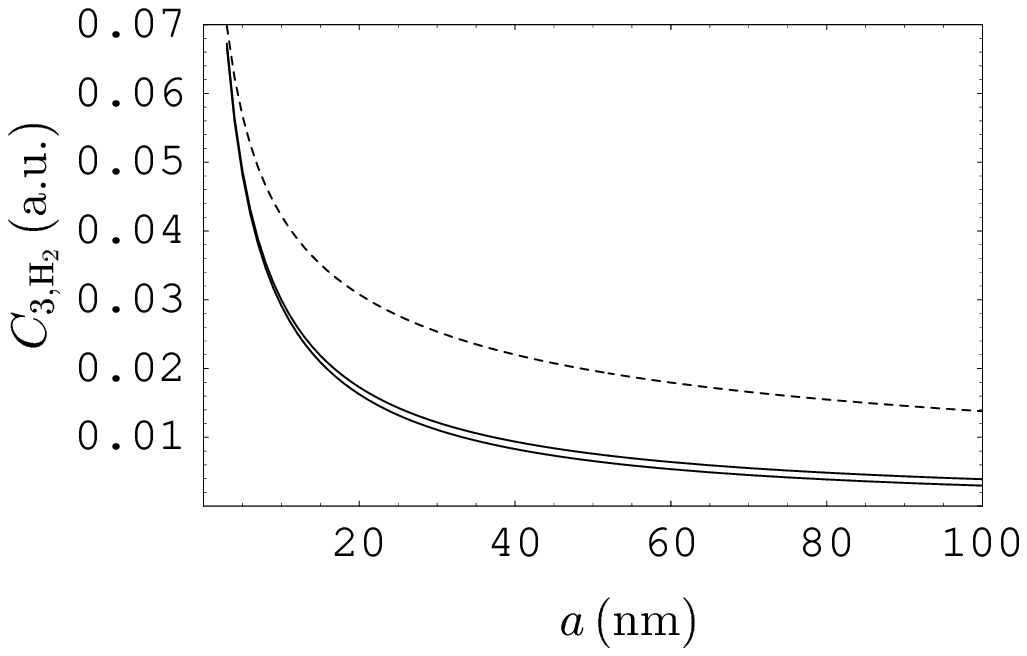}
} \vspace*{-6.cm} \caption{The van der Waals coefficient for a
H${}_2$ molecule interacting with graphene as a function of
separation. The dashed and the two solid lines are
computed using the hydrodynamic model and the Dirac model with
two different values of the gap parameter indicated in the text.}
\end{figure*}
\begin{figure*}[h]
\vspace*{-13.cm}
\centerline{\hspace*{2.5cm}
\includegraphics{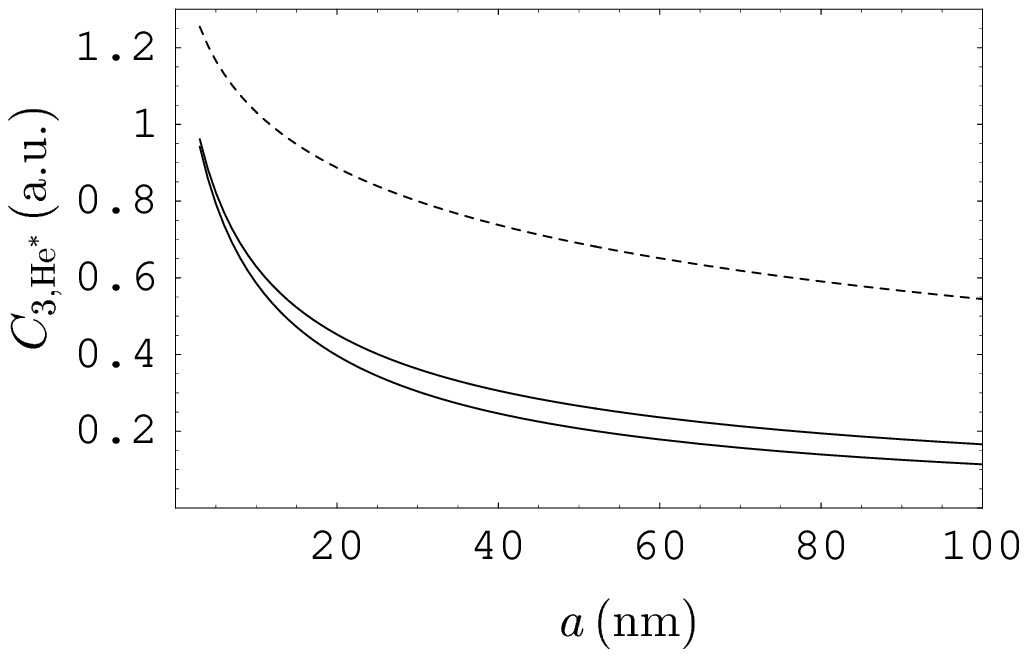}
} \vspace*{-6.cm} \caption{The van der Waals coefficient for a
He${}^{\ast}$ atom interacting with graphene as a function of
separation. The dashed and the two solid lines are
computed using the hydrodynamic model and the Dirac model with
two different values of the gap parameter indicated in the text.}
\end{figure*}
\begin{figure*}[h]
\vspace*{-13.cm}
\centerline{\hspace*{2.5cm}
\includegraphics{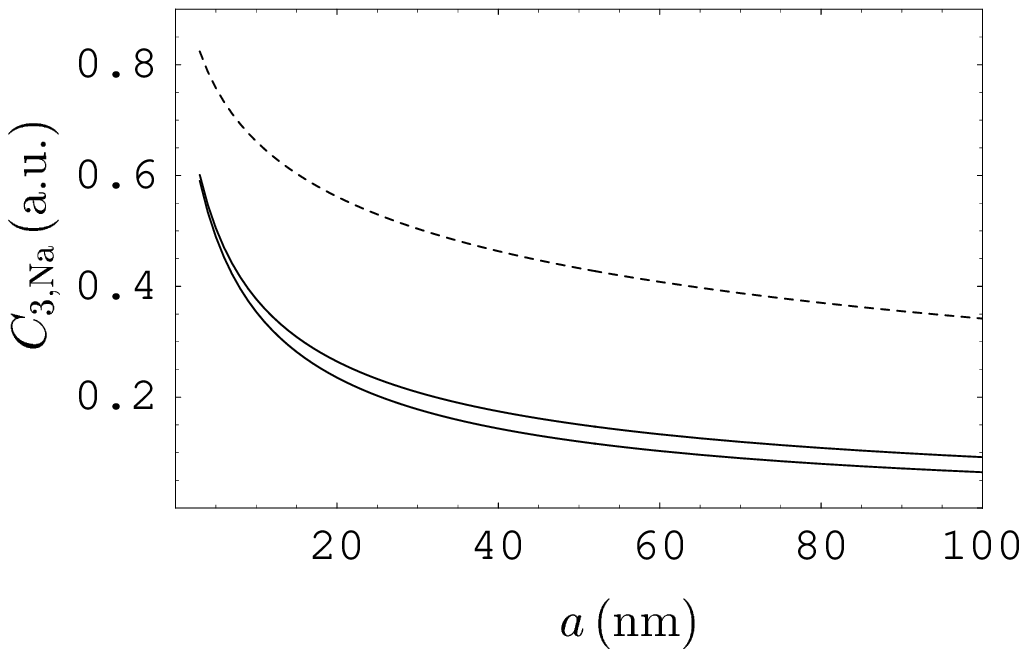}
} \vspace*{-6.cm} \caption{The van der Waals coefficient for a
Na atom interacting with graphene as a function of separation.
 The dashed and the two solid lines are
computed using the hydrodynamic model and the Dirac model with
two different values of the gap parameter indicated in the text.}
\end{figure*}
\end{document}